\documentclass[journal,11pt,draftclsnofoot,onecolumn]{IEEEtran}

\usepackage{graphicx}

\usepackage[sort,compress]{cite}
\usepackage[usenames,dvipsnames]{color}
\usepackage{latexsym}
\usepackage{psfrag}
\usepackage{amsthm,amssymb,amscd,amssymb,amsmath,mathrsfs}
\usepackage[noend]{algpseudocode}

\theoremstyle{remark}

\newtheorem{thm}{Theorem}

\newtheorem{assum}{Assumption}

\newtheorem{rem}{Remark}

\begin{document}
%
\title{Local Average Consensus in Distributed Measurement of Spatial-Temporal Varying Parameters: 1D Case}
%
%
%

\author{Kai Cai, Brian D.O. Anderson, Changbin Yu, and Guoqiang Mao 
\thanks{K. Cai is with the University of Toronto, Toronto,
Ontario M5S 3G4, Canada. His work was supported by the Australian
National University and NICTA Ltd, Canberra, ACT Australia. Email:
kai.cai@scg.utoronto.ca.  B.D.O. Anderson is with the Australian
National University and NICTA Ltd, Canberra, A.C.T., Australia.  His
work was supported by the Australian Research Council through
DP-130103610 and  DP-110100538 and by National ICT Australia. Email:
brian.anderson@anu.edu.au. C. Yu is with the Australian National
University and NICTA Ltd, Canberra, A.C.T., Australia, and also with
Shandong Computer Science Center, Jinan, China. The work of C. Yu
was supported by the Australian Research Council through
DP-130103610 and a Queen Elizabeth II Fellowship under DP-
110100538, and the Overseas Expert Program of Shandong Province.
Email: brad.yu @anu.edu.au. G. Mao is with the School of Electrical
and Information Engineering, University of Sydney and National ICT
Australia (NICTA). Email: guoqiang.mao@sydney.edu.au.
}
}

\maketitle

\begin{abstract}
We study a new variant of consensus problems, termed `local average
consensus', in networks of agents.  We consider the task of using
sensor networks to perform distributed measurement of a parameter
which has both spatial (in this paper 1D) and temporal variations.
Our idea is to maintain potentially useful local information
regarding spatial variation, as contrasted with reaching a single,
global consensus, as well as to mitigate the effect of measurement
errors. We employ two schemes for computation of local average
consensus: exponential weighting and uniform finite window.  In both
schemes, we design local average consensus algorithms to address
first the case where the measured parameter has spatial variation
but is constant in time, and then the case where the measured
parameter has both spatial and temporal variations.  Our designed
algorithms are distributed, in that information is exchanged only
among neighbors. Moreover, we analyze both spatial and temporal
frequency responses and noise propagation associated with the
algorithms.  The tradeoffs of using local consensus, as compared to
standard global consensus, include higher memory requirement and
degraded noise performance.  Arbitrary updating weights and random
spacing between sensors are analyzed in the proposed algorithms.
\end{abstract}
%


\section{Introduction}
\label{Sec_Intro}

Consensus of multi-agent systems comes in many varieties (e.g.
\cite{BerTsi:89,JadLinMor:03,Mor:05,RenBeaAtk:07,SaFaMu:07}), and in
this paper, we focus on a particular variety, namely average
consensus (e.g.
\cite{OlfMur:04,BoGhPrSh:06,XBK07,CaiIshii:AUT12,TK12}). This refers
to an arrangement where each of a network of agents is associated
with a value of a certain variable, and a process occurs which ends
up with all agents learning the average value of the variable.
Finding an average of a set of values is apparently conceptually
trivial; what makes average consensus nontrivial is the fact that an
imposed graphical structure limits the nature of the steps that can
be part of the averaging algorithm, each agent only being allowed to
exchange information with its neighbors, as defined by an overlaid
graphical structure. Issues also arise of noise performance,
transient performance, effect of time delay, agent/link loss, etc
(\cite{KarMou:09,LovZam:10,LiuLuChen:10,ZhangTian:12}).

Finding an average also throws away much information. In many
situations, one might well envisage that a \emph{local average}
might be useful, retaining the characteristics of local information
meanwhile mitigating the effect of measurement error. For instance,
one thousand weather stations across a city, instead of giving a
single air pollution reading, might validly be used to identify
hotspots of pollution, i.e. localities with high pollution; thus,
instead of a global average, a form of local averaging, still
mitigating the effects of some noise, might be useful.

We term this variant `local (average) consensus', and distinguish it
from the normal sort of consensus, termed here by way of contrast
`global (average) consensus'.

We consider two schemes for computation of local average consensus.
One involves the use of exponential weights to reflect `closeness'
of the agents measured in both topological and geographical distance
(viz. the further a neighbor is, the lesser its value will affect
the agent's computation of its `local average').  The other scheme
employs a finite window to reduce computation burden; the bounds of
the finite window will be case-dependent in applications.  In both
schemes, we design local consensus algorithms to address first the
case where the measured variable has spatial variation but is
constant in time, and then the case where the measured variable has
both spatial and temporal variations. In this paper we consider
spatial variation in 1D for simplicity. The designed local consensus
algorithms are distributed, as their global consensus counterparts,
in that information exchange is allowed only among neighbors. As we
will see, these algorithms have higher memory requirement than that
of a global consensus algorithm (the latter can be made memoryless).

We also seek to understand the properties of the designed local
consensus algorithms. In particular, we analyze both spatial and
temporal frequency responses and noise propagation associated with
the algorithms.  To obtain a fully analytical result we limit our
study to a 1D sensor network, which can find its application in
power line monitoring, canal/river monitoring, detection of border
intrusions, structural monitoring of railways/bridges/pipelines, etc
\cite{SI:03,ChenHwang:08,HYA12,ArikAkan:10,YYHLS:07}. Moreover, we
investigate two generalizations of the designed local consensus
algorithms, one with arbitrary updating weights and the other with
random spacing between sensors.

We note that \cite{OlfSha:05} proposed a ``consensus filter'' which
allows the nodes of sensor networks to track the average of their
time-varying noisy measurements. This problem is called ``dynamic
average consensus'', which is later further studied in e.g.
\cite{FreYaLyn:06,BaiFreLyn:10}, and also in
\cite{HongHuGao:06,CaoRenLi:09,BaiArcWen:09} under a different name
``coordinated average tracking''. These works, however, deal still
with global average consensus, because all nodes are required to
track \emph{the same} time-varying average value. By contrast, our
goal of local average consensus is to have each node track the
time-varying average value only within its spatial neighborhood,
thereby retaining characteristics of locally measured information.

The rest of the paper is organized as follows.
Section~\ref{Sec_Timeinv} presents local average consensus
algorithms for the case where the measured variable has spatial
variation but is constant in time.
Section~\ref{Sec_FreqResp_spatial} and Section~\ref{Sec_Noise}
investigate spatial frequency response and noise propagation of the
designed algorithms. Section~\ref{Sec_random} studies arbitrary
weights and random spacing in the proposed local averaging
algorithms. Section~\ref{Sec_Timevarying} presents local consensus
algorithms for the case where the measured variable has both spatial
and temporal variations. This allows the treatment of
Section~\ref{Sec_FreqResp_temporal} of the frequency response
associated with time variations. Finally, Section~\ref{Sec_Concl}
states our conclusions. An initial version of this paper has been
submitted for IEEE Conference on Decision and Control 2013. This
version differs from the conference predecessor through inclusion of
proofs of results, development of material on the frequency response
to time-variation in measured variables, and analysis of random
spacing and arbitrary weights in the proposed algorithms.


%
\section{Distributed Local Consensus Algorithms} \label{Sec_Timeinv}

Consider a variable whose values vary in 1D space, and/or in
addition vary in time.
Suppose we have a (possibly infinite) chain of sensors to be placed
(uniformly) along the 1D space. Each sensor $i$ has two variables: a
measurement variable $x_i$ and a consensus variable $y_i$. At each
time $k=0,1,2,...$ each sensor $i$ takes a measurement $x_i(k)$
(potentially noisy) of the variable.  Our goal is to design
distributed algorithms which update each sensor $i$'s consensus
variable $y_i(k)$, based on $x_i(k)$ and information only from the
two immediate neighbors $i-1$ and $i+1$, such that $y_i(k)$
converges to a value which reflects spatial-temporal variations of
the variable (as we define below).

In this section, we focus on the case where all local measurements
are time-invariant, i.e. $x_i(k)=x_i$ (a constant) for all $i, k$.
The time-varying case will be addressed in
Section~\ref{Sec_Timevarying}, below.  We consider two types of
weighting schemes: exponential weighting and uniform finite window.

\subsection{Exponential Weighting}
\label{Subec_ExpoWeig}

For computing a local average at sensor $i$, it is natural to assign
larger weights to information that is spatially closer to $i$.  One
way of doing so is to assign an exponential weight $\rho^j$, $\rho
\in (0,1)$ and $j$ a nonnegative integer, to a measurement taken at
distance $j$ from $i$. For this scheme, we formulate the following
problem, adopting the reasonable assumption that there is a bound $M
< \infty$ such that measurement variables $|x_i| < M$ for all $i$.

\emph{Problem 1.} Let $\rho \in (0,1)$. Design a distributed
algorithm to update each sensor $i$'s consensus variable $y_i(k)$
such that
\begin{align} \label{prob1}
\lim_{k \rightarrow\infty} y_i(k) = \frac{1-\rho}{1+\rho} \left( x_i
+ \sum_{j=1}^\infty \rho^j (x_{i-j}+x_{i+j}) \right).
\end{align}
Thus, exponentially decaying weights, at the rate $\rho$, are
assigned to the information from both forward and backward
directions.  Note that the limit of $y_i(k)$ exists because all
$x_i$ are assumed bounded. The scaling constant $(1-\rho)/(1+\rho)$
ensures that, if all $x_i$ are the same, $y_i(k)$ is in the limit
equal to $x_i$.

We propose the following distributed algorithm to solve Problem~1.
For all $i$,
\begin{subequations} \label{alg_inv_exp}
{\small \begin{align}
y_i(0) &= \frac{1-\rho}{1+\rho} x_i \label{alg_inv_expa} \tag{\theequation a}\\
y_i(1) &= y_i(0)+ \rho(y_{i-1}(0)+y_{i+1}(0)) \label{alg_inv_expb} \tag{\theequation b}\\
y_i(2) &= y_i(1)+ \rho(y_{i-1}(1)-y_{i-1}(0)) +  \label{alg_inv_expc} \tag{\theequation c}\\
&\hspace{0.5cm} \rho(y_{i+1}(1)-y_{i+1}(0)) - \rho^2 2y_i(0)  \notag\\
y_i(k+1)&=y_i(k)+ \rho(y_{i-1}(k)-y_{i-1}(k-1)) + \label{alg_inv_expd} \tag{\theequation d}\\
&\hspace{-1.2cm} \rho (y_{i+1}(k)-y_{i+1}(k-1)) - \rho^2
(y_i(k-1)-y_i(k-2)),\ k\geq 2. \notag
\end{align}}
\end{subequations}
Each sensor $i$ needs information only from its two immediate
neighbors: $y_{i-1}(k)$ and $y_{i+1}(k)$, $k=0,1,...$. At each
iteration $k\ (\geq 2)$, the quantities used to update $y_i(k)$ are
$y_{i-1}(k)-y_{i-1}(k-1)$, $y_{i+1}(k)-y_{i+1}(k-1)$, and
$y_i(k-1)-y_i(k-2)$. Thus more memory is required in this local
consensus algorithm than in a global consensus algorithm, though the
increase is obviously modest.

\begin{thm} \label{thm:alg_inv_exp}
Algorithm~(\ref{alg_inv_exp}) solves Problem~1.
\end{thm}
\emph{Proof.} We will show by induction on $k \geq 1$ that
\begin{align} \label{prob1_k}
y_i(k) = y_i(k-1) + \rho^k (y_{i-k}(0)+y_{i+k}(0)),\ \ \forall i.
\end{align}
This leads to
\begin{align*} 
y_i(k) &= y_i(0) + \sum_{j=1}^k \rho^j
(y_{i-j}(0)+y_{i+j}(0))\\
&=\frac{1-\rho}{1+\rho} \left( x_i + \sum_{j=1}^k \rho^j
(x_{i-j}+x_{i+j}) \right),\ \ \forall i.
\end{align*}
The second equality above is due to (\ref{alg_inv_expa}). Then
taking the limit as $k \rightarrow \infty$ yields (\ref{prob1}).
That the limit exists follows from the fact that $|x_i| < M <
\infty$ and $\rho \in (0,1)$.

First, it is easily verified from (\ref{alg_inv_expb}),
(\ref{alg_inv_expc}) that (\ref{prob1_k}) holds when $k=1,2$.  Now
let $k \geq 2$ and suppose (\ref{prob1_k}) holds for all
$k'\in[1,k]$. According to (\ref{alg_inv_expd}) we derive
\begin{equation} \label{proof1}
\begin{split}
y_i(k+1)&=y_i(k)+ \rho(\rho^k (y_{i-k-1}(0)+y_{i+k-1}(0))) + \\
&\hspace{0.5cm} \rho(\rho^k (y_{i-k+1}(0)+y_{i+k+1}(0))) - \\
&\hspace{0.5cm} \rho^2 (\rho^{k-1} (y_{i-k+1}(0)+y_{i+k-1}(0))) \\
&= y_i(k) + \rho^{k+1} (y_{i-k-1}(0)+y_{i+k+1}(0)).
\end{split}
\end{equation}
Therefore, (\ref{prob1_k}) holds for all $k \geq 1$. \hfill
$\blacksquare$

Note from the derivation in (\ref{proof1}) that in the scheme
(\ref{alg_inv_expd}), $y_{i-1}(k)-y_{i-1}(k-1)$ produces new
information $y_{i-k-1}(0)+y_{i+k-1}(0)$ (resp.
$y_{i+1}(k)-y_{i+1}(k-1)$ produces $y_{i-k+1}(0)+y_{i+k+1}(0)$), and
$y_i(k-1)-y_i(k-2)$ is a correction term which cancels the redundant
information $y_{i-k+1}(0)+y_{i+k-1}(0)$.

\begin{rem} \label{rem:alg_inv_exp_ext}
An extension of Algorithm~(\ref{alg_inv_exp}) is immediate.
Each sensor $i$ weights information from the backward direction
differently from the forward direction, using exponential weights
$\rho_{b}$ and $\rho_{f} \in (0,1)$, respectively. Then revise
Algorithm~(\ref{alg_inv_exp}) as follows:
\begin{subequations} \label{alg_inv_exp_ext2}
{\small \begin{align}
y_i(0) &= \frac{(1-\rho_{b})(1-\rho_{f})}{1-\rho_{b}\rho_{f}} x_i \label{alg_inv_expa_ext2} \tag{\theequation a}\\
y_i(1) &= y_i(0)+ \rho_{b}y_{i-1}(0) + \rho_{f}y_{i+1}(0) \label{alg_inv_expb_ext2} \tag{\theequation b}\\
y_i(2) &= y_i(1)+ \rho_{b}(y_{i-1}(1)-y_{i-1}(0)) +  \label{alg_inv_expc_ext2} \tag{\theequation c}\\
&\hspace{0.5cm} \rho_{f}(y_{i+1}(1)-y_{i+1}(0)) - \rho_{b}\rho_{f} 2y_i(0)  \notag\\
y_i(k+1)&=y_i(k)+ \rho_{b}(y_{i-1}(k)-y_{i-1}(k-1)) + \label{alg_inv_expd_ext2} \tag{\theequation d}\\
&\hspace{-1.2cm} \rho_{f}(y_{i+1}(k)-y_{i+1}(k-1)) -
\rho_{b}\rho_{f} (y_i(k-1)-y_i(k-2)), \notag\\
&\hspace{6.2cm} k\geq 2. \notag
\end{align}}
\end{subequations}
This revised algorithm yields {\footnotesize
\begin{align}
\label{prob1_ext2} \lim_{k \rightarrow\infty} y_i(k) =
\frac{(1-\rho_{b})(1-\rho_{f})}{1-\rho_{b}\rho_{f}} \left( x_i +
\sum_{j=1}^\infty (\rho_{b}^j x_{i-j} + \rho_{f}^j x_{i+j}) \right).
\end{align}}
The proof of this claim is almost the same as that validating
Algorithm~(\ref{alg_inv_exp}).
\end{rem}


\subsection{Uniform Finite Window}

An alternative to exponential weighting is to have a finite window
for each sensor such that every agent's information within the
window is weighted uniformly, and the information outside the window
discarded.  For time-invariant measurements, this is to compute the
average of measurements within the window.  We formulate the
problem.

\emph{Problem 2.} Let $L \geq 1$ be an integer, and $2L+1$ the
length of the finite window of sensor $i$; i.e. sensor $i$ uses
measurement information from $L$ neighbors in each direction.
Suppose $i$ knows $L$.  Design a distributed algorithm to update
each $i$'s consensus variable $y_i(k)$ such that
\begin{equation} \label{prob2}
\begin{split}
y_i(L) &= \frac{1}{2L + 1} \left( x_i + \sum_{j=1}^{L}
(x_{i-j}+x_{i+j}) \right).
\end{split}
\end{equation}
Thus it is required that the average of $2L+1$ measurements be
computed in $L$ steps.

A variation of Algorithm~(\ref{alg_inv_exp}) will solve Problem~2.
\begin{subequations} \label{alg_inv_win}
{\small \begin{align}
y_i(0) &= \frac{1}{2L + 1} x_i \label{alg_inv_wina} \tag{\theequation a}\\
y_i(1) &= y_i(0)+ (y_{i-1}(0)+y_{i+1}(0)) \label{alg_inv_winb} \tag{\theequation b}\\
y_i(2) &= y_i(1)+ (y_{i-1}(1)-y_{i-1}(0)) +  \label{alg_inv_winc} \tag{\theequation c}\\
&\hspace{0.5cm} (y_{i+1}(1)-y_{i+1}(0)) - 2y_i(0)  \notag\\
y_i(k+1)&=y_i(k)+ (y_{i-1}(k)-y_{i-1}(k-1)) + \label{alg_inv_wind} \tag{\theequation d}\\
&\hspace{-1.2cm} (y_{i+1}(k)-y_{i+1}(k-1)) - (y_i(k-1)-y_i(k-2)),
\notag\\
&\hspace{3.8cm}  k \in [2,L-1]. \notag
\end{align}}
\end{subequations}
The memory requirement of this algorithm is the same as
Algorithm~(\ref{alg_inv_exp}): i.e. $y_{i-1}(k)-y_{i-1}(k-1)$,
$y_{i+1}(k)-y_{i+1}(k-1)$, and $y_i(k-1)-y_i(k-2)$ are needed to
update $y_i(k)$ for $k \in [2,L-1]$.  Note, however, that the
present algorithm terminates after $L$ steps because of finite
window as well as static measurements.  When measurements are
time-varying (see Section~\ref{subsec_var_win} below), by contrast,
the corresponding algorithm will need to keep track of temporal
variations. Indeed, a significant variant on
Algorithm~(\ref{alg_inv_win}) is needed, while the variation
required for Algorithm~\ref{alg_inv_exp} in comparison is minor.

\begin{thm} \label{thm:alg_inv_win}
Algorithm~(\ref{alg_inv_win}) solves Problem~2.
\end{thm}
\emph{Proof.} Similar to the proof of Theorem~\ref{thm:alg_inv_exp},
we derive for $k \in [1,L]$ that
\begin{align} \label{prob2_k}
y_i(k) = y_i(k-1) + (y_{i-k}(0)+y_{i+k}(0)),\ \ \forall i.
\end{align}
This leads to
\begin{align*} 
y_i(L) &= y_i(0) + \sum_{j=1}^{L} (y_{i-j}(0)+y_{i+j}(0))\\
&=\frac{1}{2L+1} \left( x_i + \sum_{j=1}^{L} (x_{i-j}+x_{i+j})
\right),\ \ \forall i.
\end{align*}
The second equality above is due to (\ref{alg_inv_wina}).  \hfill
$\blacksquare$

\begin{rem} \label{rem:alg_inv_win_ext}
Individual sensors may have different window lengths, $L_i \geq 1$.
In this case, we impose the condition that the neighboring lengths
may differ no more than one, i.e.
\begin{align} \label{eq:Li-constraint}
|L_i - L_{i+1}| \leq 1,\ \ \ |L_i - L_{i-1}| \leq 1,\ \ \ \forall i
\end{align}
and replace $L$ by $L_i$ throughout Algorithm~(\ref{alg_inv_win}).
Then from (\ref{alg_inv_wind}) and when $k=L_i-1$ (the final
update), we have {\small
\begin{align*}
y_i(L_i)&=y_i(L_i-1)+ (y_{i-1}(L_i-1)-y_{i-1}(L_i-2)) + \\
&\hspace{-1.2cm} (y_{i+1}(L_i-1)-y_{i+1}(L_i-2)) -
(y_i(L_i-2)-y_i(L_i-3)).
\end{align*}}
Condition (\ref{eq:Li-constraint}) ensures that both
$y_{i-1}(L_i-1)$ and $y_{i+1}(L_i-1)$ exist. Hence the same argument
as that validating Algorithm~(\ref{alg_inv_win}) proves that the
revised algorithm with $L_i$ computes
\begin{align*}
y_i(L_i) =\frac{1}{2L_i+1}x_i + \sum_{j=1}^{L_i}
(\frac{1}{2L_{i-j}+1}x_{i-j}+\frac{1}{2L_{i+j}+1}x_{i+j}).
\end{align*}
\end{rem}


%
\section{Spatial Frequency Response}
\label{Sec_FreqResp_spatial}

The whole concept of local consensus is based on the precept that
global consensus may suppress too much information that might be of
interest. In effect, global (average) consensus applies a filter to
spatial information which leaves the DC component intact, and
completely suppresses all other frequencies. Our task in this
section is to study the extent to which local consensus in contrast
does not destroy all information regarding spatial variation, and
the tool we use to do this is to look at a spatial frequency
response. Further, there is a trade-off in using local consensus,
apart from additional computational complexity as noted in
Section~\ref{Sec_Timeinv}: there is less mitigation--obviously--of
the effect of noise. We also consider this point in the next
section.




We associate with the measured variable and consensus variable
sequences $\{x_i, -\infty<i<\infty\}$ and $\{y_i,-\infty<i<\infty\}$
their spatial $Z$-transforms $\mathcal X(Z), \mathcal Y(Z)$ defined
by
\begin{equation}
\mathcal X(Z)=\sum_{-\infty}^{\infty}x_iZ^{-i}\ \ \ \ \mathcal
Y(Z)=\sum_{-\infty}^{\infty}y_iZ^{-i}
\end{equation}
Spatial $Z$-transforms capture spatial frequency content, and are a
potentially useful tool for analysing the relationship between
measured variables and consensus variables.

Our aim is to understand how,  when the measured variable sequence
has spatially sinusoidal variation at frequency $\omega$, the steady
state values of the consensus variables $y_i$ depend on $\rho$ and
$\omega$. Of course, in a practical situation spatial variation may
not necessarily be sinusoidal. The benefit of the sinusoidal
analysis is that it leads to a transfer function and hence to a
concept of bandwidth for the average consensus algorithm, i.e. a
notion of a spatial frequency below which variations can be
reasonably tracked even when the algorithm is operating, while
spatially faster variations will be suppressed or filtered out in
deriving the local average consensus.
We shall first consider local consensus with exponential weighting,
and then local consensus with a uniform finite window.

\subsection{Exponential Weighting}

The calculation using $Z$-transforms proceeds as follows. Starting
with the steady state equation (cf. (\ref{prob1}))
\begin{eqnarray}
y_i=\frac{1-\rho}{1+\rho}(x_i+\rho
x_{i-1}+\rho^{2}x_{i-2}+\cdots\\\nonumber+\rho
x_{i+1}+\rho^{2}x_{i+2}+\cdots)
\end{eqnarray}
one has {\small\begin{equation} \begin{split}
Z^{-i}y_i=&\frac{1-\rho}{1+\rho}[x_iZ^{-i}+Z^{-1}\rho
x_{i-1}Z^{-(i-1)}+Z^{-2}\rho^2x_{i-2}Z^{-(i-2)}\\
&+\cdots+Z\rho x_{i+1}Z^{-(i+1)}+Z^2\rho^2x_{i+2}Z^{-(i+2)}+\cdots]
\end{split}\end{equation} }
Summing from $i=-\infty$ to $\infty$ yields {\small\begin{eqnarray*}
\mathcal
Y(Z)=\frac{1-\rho}{1+\rho}[1+Z^{-1}\rho+Z^{-2}\rho^2+\cdots+Z\rho+Z^2\rho^2+\cdots]\mathcal
X(Z)\\\nonumber =\frac{1-\rho}{1+\rho}[1+\frac{\rho Z^{-1}}{1-\rho
Z^{-1}}+\frac{\rho Z}{1-\rho Z}]\mathcal X(Z)
\end{eqnarray*}
}or
\begin{equation}\label{eq:XtoY}
\mathcal Y(Z)=\frac{(1-\rho)^2}{(1-\rho Z^{-1})(1-\rho Z)}\mathcal
X(Z)
\end{equation}
For future reference, define the transfer function
\begin{equation}\label{eq:Hdef}
\mathcal H(Z)=\frac{(1-\rho)^2}{(1-\rho Z^{-1})(1-\rho Z)}
\end{equation}
For $Z = \exp(j\omega)$, the transfer function is real and positive.
However, for arbitrary $Z$ in general its value is complex. It has
two poles which are mirror images through the unit circle of each
other.

Now suppose that the measured variable sequence $x_i$ is sinusoidal,
thus $x_i=\exp(ji\omega_0)$, where $j=\sqrt{-1}$. The associated
$Z$-transform $\mathcal X(Z)$ is formally given by
$\sum_{i=-\infty}^{\infty}x_iZ^{-i}$. When $Z=\exp(j\omega)$, there
holds
\begin{eqnarray}
\mathcal X(\exp (j\omega))=\sum_{i=-\infty}^{\infty}\exp
(ji(\omega-\omega_0)) =2\pi\delta(\omega-\omega_0)
\end{eqnarray}
where we are appealing to the fact that the delta function
$\delta(x)$  is the limit of a multiple of the Dirichlet kernel
\begin{equation}
D_N(x)=\sum_{i=-N}^N\exp(jix)=\frac{\sin((N+\frac{1}{2})x)}{\sin(x/2)}
\end{equation}
i.e.
\begin{equation}
\delta (x)=\frac{1}{2\pi} \lim_{N\rightarrow\infty}
D_N(x)=\frac{1}{2\pi} \sum_{i=-\infty}^{\infty}\exp (jix)
\end{equation}
In formal terms, it follows from (\ref{eq:XtoY}) and (\ref{eq:Hdef})
that the associated $Z$-transform of the consensus variable, i.e.
$\mathcal Y(Z)$, is given by
\begin{equation}
\mathcal Y(\exp (j\omega))=\mathcal H(\exp
(j\omega))2\pi\delta(\omega-\omega_0)
\end{equation}
Equivalently, the consensus variable is also sinusoidal at frequency
$\omega_0$ and with phase shift and amplitude defined by $\mathcal
H(\exp (j\omega_0))$. The phase shift is easily checked to be zero
for all $\omega_0$, and the amplitude is in fact the value of
$\mathcal H$ itself, viz.
\begin{equation} \label{eq:H_exp}
\mathcal H(\exp (j\omega_0))=\frac{(1-\rho)^2}{1+\rho^2-2\rho\cos
\omega_0}
\end{equation}
Observe that if $\omega_0=0$, i.e. the measured variable is a
constant or spatially invariant, then $\mathcal H(1)=1$ irrespective
of $\rho$, i.e. the consensus variable is the same constant -- as we
would expect. Observe further that for fixed $\omega_0\neq 0$, as
$\rho\rightarrow 1$, $\mathcal H(\exp(j\omega_0))\rightarrow 0$,
which is consistent with the fact that with $\rho = 1$, the average
value of the measured variable, viz. 0, will propagate through to be
the value everywhere of the consensus variable.

\begin{figure}[!t]
\centering
\includegraphics[width=0.53\textwidth]{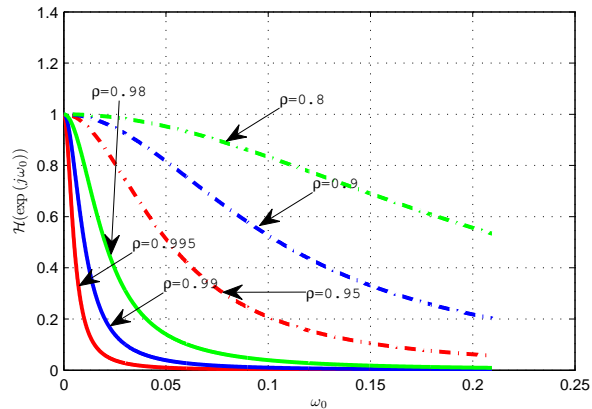}
\caption{Plot of $\mathcal H(\exp(j\omega_0))$ in (\ref{eq:H_exp})
near origin for different values of $\rho$} \label{fig:H_exp1}
\end{figure}

\begin{figure}[!t]
\centering
\includegraphics[width=0.53\textwidth]{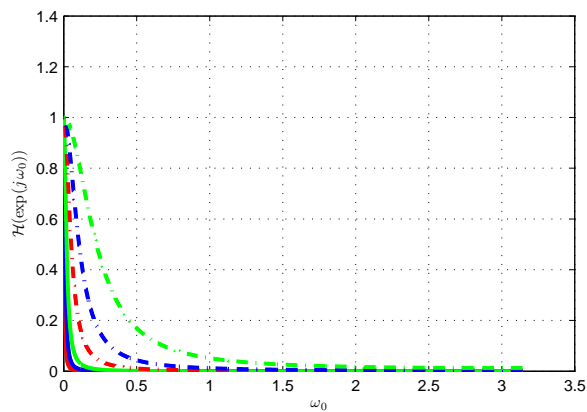}
\caption{Plot of $\mathcal H(\exp(j\omega_0))$ in (\ref{eq:H_exp})
over $[0,\pi]$ for different values of $\rho$. The colour coding is
as for Figure~\ref{fig:H_exp1}.} \label{fig:H_exp2}
\end{figure}

Observe that if $\rho$ is close to 1, i.e. $1-\rho$ is small, a
straightforward calculation shows that with $\omega_0=1-\rho$, the
value of $\mathcal H$ is approximately $1/2$. Thus crudely, $\rho$
(for values close to 1) determines the bandwidth as $O(1-\rho)$.
More generally,  we observe from the Figures~\ref{fig:H_exp1} and
\ref{fig:H_exp2} (which show behaviour near the origin and over
$[0,\pi]$), that
\begin{enumerate}
\item
For any $\rho$, $\mathcal H(\exp(j\omega_0))$ is monotonic
decreasing in $\omega_0$, from a value of 1 at $\omega_0=0$ to a
value of $\frac{(1-\rho)^2}{(1+\rho)^2}$ at $\omega_0=\pi$.
\item
For values of $1-\rho$ between zero and at least 0.2, $\mathcal
H(\exp(j\omega_0))$ takes a value of about $\frac{1}{2}$ when
$\omega_0=1-\rho$.
\end{enumerate}

The above calculations assume that there are an infinite number of
measuring agents. When the number is finite, it is clear that the
results will undergo some variation. When the hop distance to the
array boundary, call it $d$, from a particular agent, is such that
$\rho^d$ is very small, the error will obviously be minor. In the
vicinity of the boundary, the errors will be greater, and a kind of
end effect will be observed.  The results for an infinite number of
agents are accordingly indicative of the results for a finite
number.

\subsection{Uniform Finite Window}

From (\ref{prob2}), the steady-state equation in this case is
\begin{equation}
y_i=\frac{1}{2{L}+1}\sum_{k=-{L}}^{L}x_{i+k}
\end{equation}
and it is straightforward to establish that
\begin{equation}
\mathcal Y(Z)=\frac{1}{2{L}+1}\sum_{k=-{L}}^{L}Z^k\mathcal X(Z)
\end{equation}
The transfer function $\mathcal H(Z)$ is simply
$\frac{1}{2{L}+1}\sum_{k=-{L}}^{L}Z^k$ so that
\begin{equation} \label{eq:H_win}
\mathcal
H(\exp(j\omega))=\frac{1}{2{L}+1}\frac{\sin(({L}+\frac{1}{2})\omega)}{\sin(\omega/2)}
\end{equation}
The shape of the Dirichlet kernel is well known; $\mathcal H$
assumes its maximum value of 1 at $\omega=0$, and the bandwidth is
roughly $\frac{1.7}{L+1/2}$. Evidently, the bandwidths in the
exponential weighted case and the uniform finite window case are of
the same order when
\begin{align} \label{eq:equal-band}
1-\rho=\frac{1.7}{L+1/2}.
\end{align}
Put another way, and roughly speaking, a window length of $2L+1$
allows spatial variation of a bandwidth $\Omega$ to pass through the
averaging process when $L \Omega$ is about $1.7$.

\section{Noise Propagation} \label{Sec_Noise}

As mentioned already, the noise performance when local consensus is
used will be worse than that when global consensus is used. To fix
ideas, suppose that for each $i$, measurement agent $i$ has its
measurement contaminated by additive noise $\epsilon_i$ of zero mean
and variance $\sigma^2$, with the noise at any two agents being
independent.

Then if there are $N$ agents, the error in the average will be
$(1/N)\sum_{i=1}^N\epsilon_i$, which has variance
$\frac{\sigma^2}{N}$. Obviously this goes to zero as
$N\rightarrow\infty$.

When  the uniform finite window of length $2{L_i}+1$ is used, this
same thinking shows that the error variance is
$\frac{\sigma^2}{2{L_i}+1}$.

Now suppose that exponential weighting is used. In local average
consensus the error will be
\begin{equation}
\frac{1-\rho}{1+\rho}[\epsilon_i+\rho\epsilon_{i-1}+\rho^2\epsilon_{i-2}+\dots+\rho\epsilon_{i+1}+\rho^2\epsilon_{i+2}+\dots]
\end{equation}
and the variance is given by
\begin{eqnarray}
&\big(\frac{1-\rho}{1+\rho}\big)^2
[1+2\rho^2+2\rho^4+\dots]\sigma^2\\\nonumber
&~~=\big(\frac{1-\rho}{1+\rho}\big)^2[\frac{2}{1-\rho^2}-1]\sigma^2\\\nonumber
&=(1-\rho)\frac{1+\rho^2}{(1+\rho)^3}\sigma^2
\end{eqnarray}
This lies in the interval $(\frac{1}{4}(1-\rho),1-\rho))$,  and for
$\rho$ close to 1, the error is approximately equal to the lower
bound. Indeed, the closer $\rho$ is to 1, the less is the error
variance. It is not hard to verify that a uniform finite window of
length $2{L_i}+1$ and an exponential weighting of
$\rho=\frac{2{L_i}-3}{2{L_i}+1}$ yield the same variance.
Equivalently, this condition is $1-\rho=\frac{2}{{L_i}+1/2}$, which
means that exponential weighting and uniform finite window
weighting, if they achieve the same bandwidth (cf.
(\ref{eq:equal-band})), also have approximately the same noise
performance. The same condition incidentally says that
$\rho^{L_i}\approx e^{-1}$, implying that the finite window width
with uniform weighting has width determined by the number of steps
over which the exponential weighting dies off by a factor of $e$.
These observations also mean, unsurprisingly, that when ${L_i}$ or
$\rho$ are adjusted, noise variance is proportional to bandwidth.


%
\section{Generalizations} \label{Sec_random}

\subsection{Arbitrary Weighting}

To this point, we have considered two types of weights. It is at
least of academic interest to consider what might happen with
essentially \emph{arbitrary} weights. These might for example
reflect known and nonuniform spacings between agents.  We adopt the
following assumption.

\begin{assum} \label{ass:arb}
Let $a_{ij} \neq 0$ for all $i,j$. For every $i$, the following sum
\begin{align}
a_{ii}x_i + \sum_{j=1}^\infty (a_{i,i-j}x_{i-j}+a_{i,i+j}x_{i+j})
\end{align}
is finite, and
\begin{align}
K := a_{ii} + \sum_{j=1}^\infty (a_{i,i-j}+a_{i,i+j}).
\end{align}
\end{assum}

\emph{Problem 3.} Design a distributed algorithm to update each
sensor $i$'s consensus variable $y_i(k)$ such that
\begin{align} \label{prob3}
\lim_{k \rightarrow\infty} y_i(k) = \frac{1}{K} \left( a_{ii}x_i +
\sum_{j=1}^\infty (a_{i,i-j}x_{i-j}+a_{i,i+j}x_{i+j}) \right).
\end{align}
The constant $1/K$ ensures again that, if all $x_i$ are the same,
$y_i(k)$ is in the limit equal to $x_i$.

To solve Problem~3, we consider a modified approach: Let each sensor
$i$ have two additional consensus variables, $y^F_i(k)$ and
$y^B_i(k)$; $y^F_i(k)$ (resp. $y^B_i(k)$) is updated based on $x_i$
and information from the forward neighbor $i+1$ (resp. the backward
neighbor $i-1$). This approach separates the updates of consensus
variables between the forward and the backward directions. As we
will see, the separation effectively avoids term cancelations needed
in the algorithms in Section~\ref{Sec_Timeinv}, which we find
difficult in the case of arbitrary weights.


Now using the two consensus variables $y^F_i(k)$ and $y^B_i(k)$, we
present the following distributed algorithm. For all $i$,
\begin{subequations} \label{alg_inv_arb}
{\small \begin{align}
y^F_i(0) &= y^B_i(0) = \frac{1}{K} a_{ii} x_i \label{alg_inv_arba} \tag{\theequation a}\\
y^F_i(1) &=y^F_i(0) + \frac{a_{i,i+1}}{a_{i+1,i+1}}y^F_{i+1}(0) \label{alg_inv_arbb} \tag{\theequation b}\\
y^B_i(1) &=y^B_i(0) + \frac{a_{i,i-1}}{a_{i-1,i-1}}y^B_{i-1}(0) \notag\\
y^F_i(2) &=y^F_i(1) + \frac{a_{i,i+2}}{a_{i+1,i+2}} (y^F_{i+1}(1)-y^F_{i+1}(0)) \label{alg_inv_arbc} \tag{\theequation c}\\
y^B_i(2) &=y^B_i(1) + \frac{a_{i,i-2}}{a_{i-1,i-2}} (y^B_{i-1}(1)-y^B_{i-1}(0)) \notag \\
y^F_i(k+1)&=y^F_i(k)+ \frac{a_{i,i+k+1}}{a_{i+1,i+k+1}}
(y^F_{i+1}(k)-y^F_{i+1}(k-1)) \label{alg_inv_arbd} \tag{\theequation d}\\
y^B_i(k+1)&=y^B_i(k)+ \frac{a_{i,i-k-1}}{a_{i-1,i-k-1}}
(y^B_{i-1}(k)-y^B_{i-1}(k-1)),\ k\geq 2. \notag
\end{align}}
\end{subequations}
In the above algorithm, each sensor $i$ requires two consensus
variables and needs to know the weights used by its two neighbors,
in addition to the memory requirement of the algorithms in
Section~\ref{Sec_Timeinv}. Finally, values of $y^F_i(k)$ and
$y^B_i(k)$ are glued together to produce $y_i(k)$ as follows:
\begin{align} \label{alg_inv_arby}
y_i(k) = y^F_i(k) + y^B_i(k) - \frac{1}{K} a_{ii} x_i,\ \forall k
\geq 0.
\end{align}
The last term above serves to correct that the initial $(1/K) a_{ii}
x_i$ value in (\ref{alg_inv_arba}) is added twice

\begin{thm} \label{thm:alg_inv_arb}
Let Assumption~\ref{ass:arb} hold. Then
Algorithm~(\ref{alg_inv_arb})-(\ref{alg_inv_arby}) solves Problem~3.
\end{thm}
\emph{Proof.} First, we show by induction on $k \geq 1$ that for all
$i$,
\begin{align} \label{prob3_k}
y^F_i(k) = y^F_i(k-1) + \frac{1}{K} a_{i,i+k}x_{i+k}.
\end{align}
It is easily verified from (\ref{alg_inv_arbb}),
(\ref{alg_inv_arbc}) that (\ref{prob3_k}) holds when $k=1,2$.  Now
let $k \geq 2$ and suppose (\ref{prob3_k}) holds for $k$. According
to (\ref{alg_inv_arbd}) we derive
\begin{align*}
&y^F_i(k+1)=y^F_i(k) + \frac{a_{i,i+k+1}}{a_{i+1,i+k+1}} \frac{1}{K} a_{i+1,i+k+1}x_{i+k+1} \\
&\hspace{0.0cm}= y^F_i(k) + \frac{1}{K} a_{i,i+k+1}x_{i+k+1}.
\end{align*}
Therefore, (\ref{prob3_k}) holds for all $k \geq 1$, and leads to
\begin{align*}
y^F_i(k) &= y^F_i(0) + \frac{1}{K} \sum_{j=1}^k a_{i,i+j}x_{i+j}\\
&=\frac{1}{K} \left( a_{ii}x_i + \sum_{j=1}^k a_{i,i+j}x_{i+j}
\right),\ \ \forall i.
\end{align*}
The second equality above is due to (\ref{alg_inv_arba}). Similarly,
for $y^B_i(k)$, we derive
\begin{align*}
y^B_i(k) =\frac{1}{K} \left( a_{ii}x_i + \sum_{j=1}^k
a_{i,i-j}x_{i-j} \right),\ \ \forall i.
\end{align*}
Now by (\ref{alg_inv_arby}),
\begin{align*}
y_i(k) =\frac{1}{K} \left( a_{ii}x_i + \sum_{j=1}^k
(a_{i,i-j}x_{i-j} + a_{i,i+j}x_{i+j}) \right),\ \ \forall i.
\end{align*}
Then taking the limit as $k \rightarrow \infty$ yields
(\ref{prob3}). That the limit exists follows from
Assumption~\ref{ass:arb}. \hfill $\blacksquare$

\subsection{Random Spacing}

If the arbitrary weights studied in the previous subsection reflect
nonuniform distances between successive sensors, we may assume that
these distances are random, in accordance with some probability law.
Two different possibilities are that (a) they are Poisson
distributed, let us say with intensity 1 (assumed for convenience),
or (b) the inter sensor distances are uniformly distributed in an
interval $[1-\eta,1+\eta]$ where $\eta$ is known. Different physical
mechanisms could typically lead to these two situations. In the
first case, sensor distances are independent. In the second case, we
make the explicit assumption that inter sensor distances are
independent random variables.

Based on the treatment already derived for the case corresponding to
uniform spacing in Section~\ref{Subec_ExpoWeig}, where a weighting
of $\rho^d$ applies at a given sensor to the measurement passed to
it and made at a sensor $d$ units away, we suggest that the relevant
weighting to apply to the measurement collected at sensor $j$ and
used at sensor $i<j$ is, with $d_{i,i+j}$ denoting the distance
between sensors $i$ and $i+j$,
 $$\rho^{d_{i,i+1}+d_{i+1,i+2}+\dots+d_{j-1,j}}=\rho^{d_{ij}}$$

The full expression for the average consensus variable at node $i$
is then
\begin{equation}
y_i=K[x_i+\sum_{j=1}^{\infty}\rho^{d_{i,i+j}}x_{i+j}+\sum_{j=1}^{\infty}\rho^{d_{i,i-j}}x_{i-j}]
\end{equation}
Here $K$ is a normalization constant. In the sequel, we determine
$K$.

In the deterministic case (Section~\ref{Subec_ExpoWeig}), the
normalisation constant ($\frac{1-\rho}{1+\rho}$) was chosen to
ensure that if all measured variables had the same value, $a$ say,
then the average consensus variable also took the value $a$. In the
random case, we can seek this requirement. But it turns out that we
can only assure that $E[y_i]=a$. It would then be relevant to
consider the question of the variance in $y_i$. This is also covered
below.

Let us now assume $a=1$ for convenience. Then

\begin{equation}
y_i=K[1+\sum_{j=1}^{\infty}\rho^{d_{i,i+j}}+\sum_{j=1}^{\infty}\rho^{d_{i,i-j}}]
\end{equation}

Define two random variables
\begin{equation}
u=\sum_{j=0}^{\infty}\rho^{d_{i,i+j}},
\;\;v=\sum_{j=0}^{\infty}\rho^{d_{i,i-j}}
\end{equation}
(Take $d_{i,i}=0$, so that the first summand in each case is $1$.)
Then $u,v$ have the same distribution and are independent. It is
obvious that
\begin{equation}\label{eq:rv}
y_i=K[u+v-1]
\end{equation}
This equation makes clear that $y_i$ is indeed a random variable, so
that $K$ can only be chosen to ensure that $E[y_i]=1$. Now observe
further that
\begin{equation}\label{eq:rv2}
u=1+\rho^{d_{i,i+1}}\sum_{j=1}^{\infty}\rho^{d_{i+1,i+j}}=1+\rho^{d_{i,i+1}}w
\end{equation}
where, crucially,  $w$ evidently has the same distribution as $u$,
but is independent of the random variable $\rho^{d_{i,i+1}}$. Hence
there holds
\begin{equation}
E[u]=1+E[\rho^{d_{i,i+1}}]E[u]
\end{equation}
whence $E[u]=(1-E[\rho^{d_{i,i+1}}])^{-1}$ and then to assure
$E[y_i]=1$, equation (\ref{eq:rv}) implies that we need
\begin{equation}
K=\frac{1-E[\rho^{d_{i,i+1}}]}{1+E[\rho^{d_{i,i+1}}]}
\end{equation}

Now suppose the distribution of $d_{i,i+1}$ is Poisson with
intensity 1, for which the probability density is $e^{-d}$. The
expected value of $\rho^{d_{i,i+1}}$ is then easily computed to be
$[1-\log \rho]^{-1}$, so that
\begin{equation}
K=\frac{{\color{blue}-}\log \rho}{2-\log\rho}
\end{equation}

We remark that when $1-\rho$ is small, both $K$ and the expression
applicable in the deterministic case, viz. $\frac{1-\rho}{1+\rho}$,
are approximately $\frac{1}{2}(1-\rho)$.

If the distribution of $d_{i,i+1}$ is uniform in $[1-\eta,1+\eta]$,
then the expected value of $\rho^{d_{i,i+1}}$ is
$\frac{1}{2\eta\log\rho}[\rho^{1+\eta}-\rho^{1-\eta}]$, (the limit
of which is $\rho$ when $\eta\rightarrow 0$, as expected). The value
of $K$ in this case is
\begin{equation}
K=\frac{2\eta\log\rho-(\rho^{1+\eta}-\rho^{1-\eta})}{2\eta\log\rho+\rho^{1+\eta}-\rho^{1-\eta}}.
\end{equation}
Once again, one can verify that when $1-\rho$ is small, the
expression is approximately $\frac{1}{2}(1-\rho)$.


Now since we can only assure in the event all $x_i$ assume the value
that $E[y_i]$ takes that value, rather than $y_i$ itself, it is of
interest to consider what the error might be. Guidance as to the
error follows from the variance $E(y_i-E[y_i])^2$. We can work out
the variance also, in the following way. From (\ref{eq:rv}) and the
fact that $u,v$ are independent but with the same distribution,
there follows, in obvious notation
\begin{equation}
\sigma_y^2=2K^2\sigma_u^2
\end{equation}
Now if $x,y$ are two independent random variables with $z=xy$, there
holds
$\sigma_z^2=\sigma_x^2\sigma_y^2+\sigma_x^2E[y]^2+E[x]^2\sigma_y^2$,
and using this it follows from (\ref{eq:rv2}) and the fact that
$\xi:=\rho^{d_{i,i+1}}$ and  $w$ are independent, $w$ having the
same distribution as $u$, that
\begin{equation}
\sigma_u^2=\sigma_{\xi}^2\sigma_u^2+\sigma_{\xi}^2E[u]^2+E[\xi]^2\sigma_u^2
\end{equation}
or
\begin{equation}
\sigma_u^2=\frac{\sigma_{\xi}^2E[u]^2}{1-\sigma_{\xi}^2-E[\xi]^2}=\frac{\sigma_{\xi}^2E[u]^2}{1-E[\xi^2]}
\end{equation}
It is straightforward to check that
\begin{eqnarray}
E[\xi^2]&=&\frac{1}{1-2\log\rho}\\\nonumber
\sigma_{\xi}^2&=&\frac{1}{1-2\log\rho}-\frac{1}{(1-\log\rho)^2}\\\nonumber
\sigma_u^2&=&-\frac{1}{2\log\rho}\\\nonumber
\sigma_y^2&=&2K^2\sigma_u^2=-\frac{\log\rho}{(2-\log\rho)^2}
\end{eqnarray}

which is of the order of $-\log\rho$. When $x:=1-\rho$, this is
approximately $x$. Comparing this variance with the error variance
arising in $y_i$ with deterministic spacing but error variance
$\sigma^2=1$ of additive noise perturbing each measured variable, we
see that the error is of a similar magnitude.


%
\section{Local Consensus with Time-Varying Measurements} \label{Sec_Timevarying}

We have so far considered time-invariant local measurements.  In
practice, however, most measured variables are time-varying: e.g.
temperature, pollution, and current/voltage in power lines. In this
section, we consider that each measurement variable $x_i(k)$ is
time-varying, i.e. a function of time $k$, and design distributed
algorithms to track temporal variations of measurements, in addition
to spatial variations.

Note that in typical studies of global average consensus, it is not
common to postulate that local variables change over time.
Nevertheless, convergence rates are often considered, being
identified as exponential, and there are numerous results that seek
to identify such rates (see e.g. \cite{XiBo:04,SaFaMu:07}). The
rates themselves are indicative of the bandwidth of variation of
measured variables whose average can be tracked by the global
consensus algorithms.

In the sequel, we will again consider the two schemes: first
exponential weighting, and then uniform finite window.

\subsection{Exponential Weighting} \label{subsec_var_exp}

Henceforth, we shall assume as is reasonable that there is a bound
$M < \infty$ such that measured variables $|x_i(k)| < M$ for all
$i,k$.

\emph{Problem 3.} Let $\rho \in (0,1)$. Design a distributed
algorithm to update each sensor $i$'s consensus variable $y_i(k)$
such that
\begin{align} \label{prob1_temporal}
y_i(k) = \frac{1-\rho}{1+\rho} \Big( x_i(k) + \sum_{j=1}^k \rho^j
(x_{i-j}(k-j)+x_{i+j}(k-j)) \Big).
\end{align}
Here an exponential weight $\rho^j$ is applied to measurements from
$j$ steps away sensors in both directions with $j$ time delay.  In
this way temporal changes of $x_i$ are taken into account.

Extending Algorithm~(\ref{alg_inv_exp}), we propose the following
distributed algorithm, which differs from (\ref{alg_inv_exp}) by
inclusion of additional terms reflecting temporal changes in local
measurement values.
\begin{subequations} \label{alg_var_exp}
{\small\begin{align}
y_i(0) &= \lambda x_i(0),\ \ \lambda:=\frac{1-\rho}{1+\rho} \label{alg_var_expa} \tag{\theequation a}\\
y_i(1) &= y_i(0)+ \rho(y_{i-1}(0)+y_{i+1}(0)) + \lambda(x_i(1)-x_i(0)) \label{alg_var_expb} \tag{\theequation b}\\
y_i(2) &= y_i(1)+ \rho(y_{i-1}(1)-y_{i-1}(0)) +  \label{alg_var_expc} \tag{\theequation c}\\
&\hspace{-0.4cm} \rho(y_{i+1}(1)-y_{i+1}(0)) - \rho^2 2y_i(0) + \lambda(x_i(2)-x_i(1))  \notag\\
y_i(k+1)&=y_i(k)+ \rho(y_{i-1}(k)-y_{i-1}(k-1)) + \label{alg_var_expd} \tag{\theequation d}\\
&\hspace{-1.2cm} \rho (y_{i+1}(k)-y_{i+1}(k-1)) - \rho^2
(y_i(k-1)-y_i(k-2)) + \notag\\
&\hspace{-1.2cm} \lambda(x_i(k+1)-x_i(k)) - \rho^2 \lambda
(x_i(k-1)-x_i(k-2)),\ k\geq 2. \notag
\end{align}}
\end{subequations}
Each sensor $i$ needs information only from its two immediate
neighbors: $y_{i-1}(k)$ and $y_{i+1}(k)$, $k=0,1,...$.  Note that
sensor $i$ does not need its neighbors' measurement variables
$x_{i-1}(k)$ and $x_{i+1}(k)$. Compared to
Algorithm~(\ref{alg_inv_exp}), two additional quantities (requiring
further modest increase in local memory) are used to update
$y_i(k)$: $x_i(k+1)-x_i(k)$ and $x_i(k-1)-x_i(k-2)$; both represent
changes in local measurements at different times. As we will see
below, $x_i(k+1)-x_i(k)$ provides new information, while
$x_i(k-1)-x_i(k-2)$ is used as a correction term.

\begin{thm} \label{thm:alg_var_exp}
Algorithm~(\ref{alg_var_exp}) solves Problem~3.
\end{thm}
\emph{Proof.} It is easily verified from (\ref{alg_var_expb}) that
$y_i(1)=\lambda(x_i(1)+\rho(x_{i-1}(0)+x_{i+1}(0)))$ and from
(\ref{alg_var_expc}) that \addtocounter{equation}{1}
\begin{align} \label{proof2_a}
y_i(2) &= y_i(1)+ \rho^2(y_{i-2}(0)+y_{i+2}(0)) + \lambda \Big[ (x_i(2)-x_i(1)) \notag\\
&\hspace{-0.6cm} + \rho \big(
(x_{i-1}(1)-x_{i-1}(0))+(x_{i+1}(1)-x_{i+1}(0)) \big) \Big]
\tag{\theequation a} \\
& = \lambda \Big( x_i(2)+\rho(x_{i-1}(1)+x_{i+1}(1)) \notag\\
&\hspace{0.7cm} + \rho^2(x_{i-2}(0)+x_{i+2}(0)) \Big)
\label{proof2_b} \tag{\theequation b}
\end{align}
By (\ref{proof2_a}) we obtain the expressions of
$y_{i-1}(2)-y_{i-1}(1)$ and $y_{i+1}(2)-y_{i+1}(1)$; also by
(\ref{alg_var_expb}) we have $y_i(1)-y_i(0)$.  Substituting these
three terms into (\ref{alg_var_expd}) yields
\begin{equation} \label{proof2}
\begin{split}
y_i(3)&=y_i(2)+ \rho^3(y_{i-3}(0)+y_{i+1}(0)) + \rho^2 \lambda ((x_{i-2}(1)- \\
&\hspace{-0.5cm} x_{i-2}(0)) + (x_{i}(1)-x_{i}(0))) + \rho \lambda (x_{i-1}(2)-x_{i-1}(1)) + \\
&\hspace{-0.5cm} \rho^3(y_{i-1}(0)+y_{i+3}(0)) + \rho^2 \lambda ((x_{i}(1)-x_{i}(0)) + \\
&\hspace{-0.5cm} (x_{i+2}(1)-x_{i+2}(0))) + \rho \lambda (x_{i+1}(2)-x_{i+1}(1)) - \\
&\hspace{-0.5cm} (\rho^3 ((y_{i-1}(0)+y_{i+1}(0))) + \rho^2 \lambda (x_{i}(1)-x_{i}(0))) + \\
&\hspace{-0.5cm} \lambda(x_{i}(3)-x_{i}(2)) - \rho^2 \lambda (x_{i}(1)-x_{i}(0)) \\
&= y_i(2) + \rho^{3} (y_{i-3}(0)+y_{i+3}(0)) + \lambda \Big[ (x_i(3)-x_i(2)) \\
&\hspace{-0.5cm} + \rho \big(
(x_{i-1}(2)-x_{i-1}(1))+(x_{i+1}(2)-x_{i+1}(1)) \big) + \\
&\hspace{-0.2cm} \rho^2 \big(
(x_{i-2}(1)-x_{i-2}(0))+(x_{i+2}(1)-x_{i+2}(0)) \big) \Big].
\end{split}
\end{equation}
In deriving the second equality above, the terms $\rho^3
((y_{i-1}(0)+y_{i+1}(0)))$ and $2 \rho^2 \lambda
(x_{i}(1)-x_{i}(0))$ are canceled.  Now substituting the expression
(\ref{proof2_b}) of $y_i(2)$ into (\ref{proof2}), and canceling the
terms $\lambda x_i(2)$, $\rho \lambda (x_{i-1}(1)+x_{i+1}(1))$, and
$\rho^2 \lambda (x_{i-2}(0)+x_{i+2}(0))$, we derive
\begin{align*} 
&y_i(3) = \lambda \Big( x_i(3)+\rho(x_{i-1}(2)+x_{i+1}(2)) +\\
&\rho^2(x_{i-2}(1)+x_{i+2}(1)) +\rho^3(x_{i-3}(0)+x_{i+3}(0)) \Big).
\end{align*}
By the same procedure, inductively we can derive $y_i(k)$ for
$k=4,5,...$, and conclude that (\ref{prob1_temporal}) holds for all
$k$. \hfill $\blacksquare$

As commented in Remark~\ref{rem:alg_inv_exp_ext} for
Algorithm~(\ref{alg_inv_exp}), we may similarly extend
Algorithm~(\ref{alg_var_exp}) to the case where sensors assign
different exponential weights to information from the backward and
the forward directions, using $\rho_{b}, \rho_{f} \in (0,1)$.


\subsection{Uniform Finite Window} \label{subsec_var_win}

The finite window case with time-varying measurements is
challenging, because all information outside the window has to be
discarded, and temporal variations of information within the window
have to be tracked. We state the problem formally.

\emph{Problem 4.} Let $L \geq 1$ be an integer, and $2L+1$ the
length of the finite window of sensor $i$; i.e. sensor $i$ uses
measurement information from $L$ neighbors in each direction.
Suppose $i$ knows $L$.  Design a distributed algorithm to update
each $i$'s consensus variable $y_i(k)$ such that
\begin{equation} \label{prob2_temporal}
\begin{split}
y_i(k) &= \frac{1}{2L + 1} \left( x_i(k) + \sum_{j=1}^{k}
(x_{i-j}(k-j)+x_{i+j}(k-j)) \right)\\
& \hspace{5cm} \mbox{if $k \leq L$;} \\
y_i(k) &= \frac{1}{2L + 1} \left( x_i(k) + \sum_{j=1}^{L}
(x_{i-j}(k-j)+x_{i+j}(k-j)) \right)\\
& \hspace{5cm} \mbox{if $k > L$.}
\end{split}
\end{equation}
The explanation for the time arguments associated with $x_{i-j}$ and
$x_{i+j}$ on the right of (\ref{prob2_temporal}) is as follows. At
each time step, values can be `passed' by exactly one hop. Hence, it
takes $j$ time instances for a measured variable at sensor $i-j$ to
be perceived at sensor $j$. Therefore the consensus variable
$y_i(k)$ can depend on $x_{i-j}(k-j)$ (resp. $x_{i+j}(k-j)$ but no
later value of $x_{i-j}(k-j)$ (resp. $x_{i+j}(k-j)$).

The distributed algorithm we design to solve Problem~4 has several
features. First, it needs an additional vector of variables $z_i =
[z_{i0} \ z_{i1} \ \cdots \ z_{i(L)}]^T$ of $L+1$ components for
each sensor $i$, and $z_i$ needs to be updated along with consensus
variable $y_i$ and communicated to the two immediate neighbors $i-1$
and $i+1$. Second, the scheme for each component of $z_i$ is similar
to Algorithm~(\ref{alg_inv_win}). Finally, we will see that the
$j$th component $z_{ij}$, $j \in [0,L]$, contributes to tracking all
local measurements $x_l(k)$, $l\in[i-L,i+L]$, in the finite window
for time $k = j\ (\mbox{mod } L+1)$.

We first present the update scheme for vector $z_i$ (c.f.
Algorithm~(\ref{alg_inv_win})). For every $j \in [0,L]$, if $k < j$,
\begin{align} \label{alg_var_win_init}
z_{ij}(k) = 0;
\end{align}

\noindent if $k \geq j$ and $k = j\ (\mbox{mod } L+1)$,
\begin{subequations} \label{alg_var_winz}{\small
\begin{align}
z_{ij}(k) &= \frac{1}{2L + 1} x_i(k), \label{alg_var_winza} \tag{\theequation a}\\
z_{ij}(k+1) &= z_{ij}(k) + (z_{(i-1)j}(k)+z_{(i+1)j}(k)) \label{alg_var_winzb} \tag{\theequation b}\\
z_{ij}(k+2) &= z_{ij}(k+1)+ (z_{(i-1)j}(k+1)-z_{(i-1)j}(k))  \label{alg_var_winzc} \tag{\theequation c}\\
&\hspace{0.3cm} +(z_{(i+1)j}(k+1)-z_{(i+1)j}(k)) - 2z_{ij}(k)  \notag\\
z_{ij}(k+3)&=z_{ij}(k+2)+ (z_{(i-1)j}(k+2)-z_{(i-1)j}(k+1)) \label{alg_var_winzd} \tag{\theequation d}\\
&\hspace{-1.3cm} +(z_{(i+1)j}(k+2)-z_{(i+1)j}(k+1)) -
(z_{ij}(k+1)-z_{ij}(k)) \notag\\
&\ \ \vdots \tag{\theequation e}\\
z_{ij}(k+L)&=z_{ij}(k+L-1)+ \label{alg_var_winzf} \tag{\theequation f}\\
&\hspace{-1.cm}  (z_{(i-1)j}(k+L-1)-z_{(i-1)j}(k+L-2)) + \notag\\
&\hspace{-1.cm} (z_{(i+1)j}(k+L-1)-z_{(i+1)j}(k+L-2)) - \notag\\
&\hspace{-1.cm} (z_{ij}(k+L-2)-z_{ij}(k+L-3)) \notag
\end{align}}
\end{subequations}
The update of each component $z_{ij}$, $j \in [0,L]$, is
\emph{periodic} with period $L+1$ for $k \geq j$. The following is
the update scheme for consensus variable $y_i$.
\begin{equation} \label{alg_var_win_y}
\begin{split}
y_i(k) &= z_{ij}(k) + \sum_{l = 0, l \neq j}^{L} (z_{il}(k)-z_{il}(k-1)),\\
&\hspace{3.5cm} j = k\ (\mbox{mod } L+1).
\end{split}
\end{equation}

\emph{Example.} We provide an example to explain the above
algorithm. Let $L=2$. Then the vector $z_i = [z_{i0} \ z_{i1} \
z_{i2}]^T$, for all $i$. At $k=0$, the first variable $z_{i0}$ is
used to record the current measurement $x_i(0)$:
\begin{align*}
z_{i0}(0) &= \frac{1}{2L + 1} x_i(0) \ \ \ \mbox{by (\ref{alg_var_winza})}\\
z_{i1}(0) &= z_{i2}(0) = 0 \ \ \ \mbox{by (\ref{alg_var_win_init})} \\
y_i(0) &= z_{i0}(0) = \frac{1}{2L + 1} x_i(0) \ \ \ \mbox{by
(\ref{alg_var_win_y})}
\end{align*}
At $k=1$, $z_{i0}$ fetches measurements at $k=0$ from 1-hop
neighbors, and meanwhile the second variable $z_{i1}$ is used to
record the current measurement $x_i(1)$:
\begin{align*}
z_{i0}(1) &= z_{i0}(0) + (z_{(i-1)0}(0) + z_{(i+1)0}(0)) \ \ \ \mbox{by (\ref{alg_var_winzb})}\\
&= z_{i0}(0) + \frac{1}{2L + 1} (x_{i-1}(0)+x_{i+1}(0))\\
z_{i1}(1) &= \frac{1}{2L + 1} x_i(1) \ \ \ \mbox{by (\ref{alg_var_winza})} \\
z_{i2}(0) &= 0 \ \ \ \mbox{by (\ref{alg_var_win_init})} \\
y_i(1) &= z_{i1}(1) + (z_{i0}(1) - z_{i0}(0)) \ \ \ \mbox{by
(\ref{alg_var_win_y})}\\
&= \frac{1}{2L + 1} \left( x_i(1) + (x_{i-1}(0)+x_{i+1}(0)) \right)
\end{align*}
At $k=2$, $z_{i0}$ fetches measurements at $k=0$ from 2-hop
neighbors, $z_{i1}$ fetches measurements at $k=1$ from 1-hop
neighbors, and meanwhile the third variable $z_{i2}$ is used to
record the current measurement $x_i(2)$: {\small \begin{align*}
z_{i0}(2) &= z_{i0}(1) + (z_{(i-1)0}(1)-z_{(i-1)0}(0))+ \\
&\hspace{0.4cm} (z_{(i+1)0}(1)-z_{(i+1)0}(0)) - 2z_{i0}(0)  \ \ \ \mbox{by (\ref{alg_var_winzc})}\\
&= z_{i0}(1) + (z_{(i-2)0}(0)+z_{(i+2)0}(0))\\
&= z_{i0}(1) + \frac{1}{2L + 1} (x_{i-2}(0)+x_{i+2}(0))\\
z_{i1}(2) &= z_{i1}(1) + (z_{(i-1)1}(1) + z_{(i+1)1}(1)) \ \ \ \mbox{by (\ref{alg_var_winzb})} \\
&= z_{i1}(1) + \frac{1}{2L + 1} (x_{i-1}(1)+x_{i+1}(1))\\
z_{i2}(2) &= \frac{1}{2L + 1} x_i(2) \ \ \ \mbox{by (\ref{alg_var_winza})} \\
y_i(2) &= z_{i2}(2) + (z_{i1}(2) - z_{i1}(1)) + (z_{i0}(2) -
z_{i0}(1)) \ \ \ \mbox{by
(\ref{alg_var_win_y})}\\
&= \frac{1}{2L + 1} ( x_i(2) + (x_{i-1}(1)+x_{i+1}(1)) + \\
&\hspace{1.5cm} (x_{i-2}(0)+x_{i+2}(0)) )
\end{align*}}
Since $L=2$, information is discarded beyond 2-hop neighbors that
are outside of the finite window. Therefore the first variable
$z_{i0}$ has completed its first update cycle for measurements made
at $k=0$. Now at $k=3$, a new measurement $x_i(3)$ is made, and
$z_{i0}$ is set to record this current value. The second variable
$z_{i1}$ continues to fetch measurements at $k=1$ from 2-hop
neighbors, and $z_{i2}$ fetches measurements at $k=2$ from 1-hop
neighbors: {\small \begin{align*}
z_{i0}(3) &= \frac{1}{2L + 1} x_i(3) \ \ \ \mbox{by (\ref{alg_var_winza})} \\
z_{i1}(3) &= z_{i1}(2) + (z_{(i-1)1}(2)-z_{(i-1)1}(1))+ \\
&\hspace{0.4cm} (z_{(i+1)1}(2)-z_{(i+1)1}(1)) - 2z_{i1}(1)  \ \ \ \mbox{by (\ref{alg_var_winzc})}\\
&= z_{i1}(2) + (z_{(i-2)1}(1)+z_{(i+2)1}(1))\\
&= z_{i1}(2) + \frac{1}{2L + 1} (x_{i-2}(1)+x_{i+2}(1))\\
z_{i2}(3) &= z_{i2}(2) + (z_{(i-1)2}(2) + z_{(i+1)2}(2)) \ \ \ \mbox{by (\ref{alg_var_winzb})} \\
&= z_{i2}(2) + \frac{1}{2L + 1} (x_{i-1}(2)+x_{i+1}(2))\\
y_i(3) &= z_{i0}(3) + (z_{i2}(3) - z_{i2}(2)) + (z_{i1}(3) -
z_{i1}(2)) \ \ \ \mbox{by
(\ref{alg_var_win_y})}\\
&= \frac{1}{2L + 1} ( x_i(3) + (x_{i-1}(2)+x_{i+1}(2)) + \\
&\hspace{1.5cm} (x_{i-2}(1)+x_{i+2}(1)) )
\end{align*}}
The updates continue in the fashion that each of the three variables
$z_{i0}, z_{i1}, z_{i2}$ executes Algorithm~(\ref{alg_inv_win}) once
in each period of $3 \ (=L+1)$ time instants:
\begin{align*}
& z_{i0}:\ \ k \in [0,2],\ [3,5],\ [6,8],\ \cdots\\
& z_{i1}:\ \ k \in [1,3],\ [4,6],\ [7,9],\ \cdots\\
& z_{i2}:\ \ k \in [2,4],\ [5,7],\ [8,10],\ \cdots
\end{align*}
These update cycles are so aligned that each measurement made at a
time is taken care by exactly one variable. Note that the updates of
each variable is \emph{independent}, in the sense that the value of
one variable does not affect the update of another variable.

We now state the main result of this subsection.

\begin{thm} \label{thm:alg_var_win}
Algorithm~(\ref{alg_var_win_init})-(\ref{alg_var_win_y}) solves
Problem~4.
\end{thm}
\emph{Proof.} First, at $k=0$, we have from
(\ref{alg_var_win_init}), (\ref{alg_var_winza}) that
$z_{i0}(0)=(1/(2L + 1)) x_i(0)$ and $z_{ij}(0)=0$, $j=1,...,L$. So
by (\ref{alg_var_win_y}) $y_i(0)=z_{i0}(0)=(1/(2L + 1)) x_i(0)$.

Let $k \geq 1$ and fix $j=k\ (\mbox{mod } L+1)$. Similar to the
proof of Theorem~\ref{thm:alg_inv_win}, in particular
Equation~(\ref{prob2_k}), we derive {\small
\begin{align*}
z_{ij}(k) &= \frac{1}{2L + 1} x_i(k) \ \ \ \ \ (\mbox{again by (\ref{alg_var_winza})}) \\
z_{i(j-1)}(k) &= z_{i(j-1)}(k-1) + (z_{(i-1)(j-1)}(k-1) + \\
&\hspace{2.8cm}  z_{(i+1)(j-1)}(k-1)) \\
&= z_{i(j-1)}(k-1) + \frac{1}{2L + 1}(x_{i-1}(k-1) + x_{i+1}(k-1)) \\
z_{i(j-2)}(k) &= z_{i(j-2)}(k-1) + (z_{(i-2)(j-2)}(k-2) + \\
&\hspace{2.8cm}  z_{(i+2)(j-2)}(k-2)) \\
&= z_{i(j-2)}(k-1) + \frac{1}{2L + 1}(x_{i-2}(k-2) + x_{i+2}(k-2)) \\
&\ \ \vdots \\
z_{i0}(k) &= z_{i0}(k-1) + (z_{(i-j)0}(k-j) + z_{(i+j)0}(k-j))\\
&= z_{i0}(k-1) + \frac{1}{2L + 1}(x_{i-j}(k-j) + x_{i+j}(k-j)).
\end{align*}}

Now if $k \leq L$ (thus $j=k$), then by (\ref{alg_var_win_init})
$z_{i(j+1)}(k)=\cdots=z_{i(L)}(k)=0$.  Therefore by
(\ref{alg_var_win_y}),
\begin{align*}
y_i(k) &= z_{ij}(k) + \sum_{l=0}^{j-1} (z_{il}(k)-z_{il}(k-1))\\
&= \frac{1}{2L + 1} \left( x_i(k) + \sum_{j=1}^{k}
(x_{i-j}(k-j)+x_{i+j}(k-j)) \right).
\end{align*}
This is the first part of (\ref{prob2_temporal}).

If $k > L$, then again similar to Equation~(\ref{prob2_k}) we derive
{\small \begin{align*}
z_{i(L)}(k) &= z_{i(L)}(k-1) + (z_{(i-j-1)(L)}(k-j-1) + \\
&\hspace{2.7cm}  z_{(i+j+1)(L)}(k-j-1))\\
&= z_{i(L)}(k-1) + \frac{1}{2L + 1}(x_{i-j-1}(k-j-1) + \\
&\hspace{3.8cm} x_{i+j+1}(k-j-1))\\
&\ \ \vdots \\
z_{i(j+1)}(k) &= z_{i(j+1)}(k-1) + (z_{(i-L)(j+1)}(k-L) + \\
&\hspace{2.9cm}  z_{(i+L)(j+1)}(k-L))\\
&= z_{i(j+1)}(k-1) + \frac{1}{2L + 1}(x_{i-L}(k-L) + \\
&\hspace{4.0cm}  x_{i+L}(k-L)).
\end{align*}}
Therefore by (\ref{alg_var_win_y}),
\begin{align*}
y_i(k) &= z_{ij}(k) + \sum_{l=0, l\neq j}^{L} (z_{il}(k)-z_{il}(k-1))\\
&= \frac{1}{2L + 1} \left( x_i(k) + \sum_{j=1}^{L}
(x_{i-j}(k-j)+x_{i+j}(k-j)) \right).
\end{align*}
This is the second part of (\ref{prob2_temporal}), and thus
completes the proof.

\hfill $\blacksquare$

In the next section, we analyze the frequency response for the two
local consensus algorithms designed in this section, with respect to
both spatial and temporal variations.


%
\section{Temporal Frequency Response}
\label{Sec_FreqResp_temporal}

In this section, we consider the question of how changes in the
measured variables propagate to become changes in the consensus
variables.
Specifically, we consider how sinusoidal variations in measured
variables reflects through, as a function of frequency, to
time-variation of the local consensus variables. As with the case of
spatial variation, we are interested in understanding what speed of
variations might be trackable by the local average consensus
algorithm, through the identification of a transfer function and its
associated bandwidth.
This question is rather understudied for global consensus.
We shall first consider a special situation, viz. one where there is
no spatial variation, but merely sinusoidal time-variation, i.e. for
all $i$, there holds $x_i(k)=e^{j\omega_0 k}$. Recall that in
studying spatial variation, we considered the special case where
there was no time-variation. Studying these special situations allow
clearer examination of the separate effects of time-variation and
spatial variation.

\begin{figure}[!t]
\centering
\includegraphics[width=0.53\textwidth]{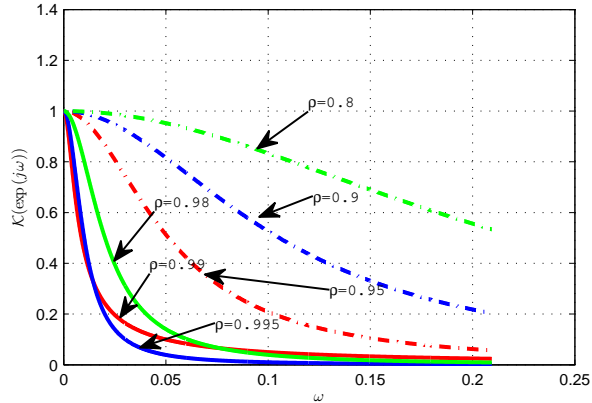}
\caption{Plot of $\mathcal K(\exp(j\omega))$ in (\ref{eq:K_exp})
near origin for different values of $\rho$} \label{fig:K_exp1}
\end{figure}

\begin{figure}[!t]
\centering
\includegraphics[width=0.53\textwidth]{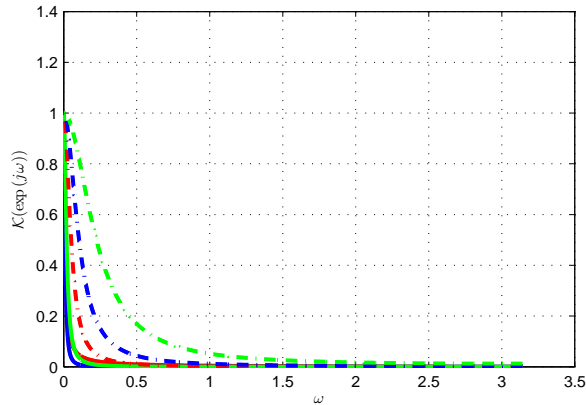}
\caption{Plot of $\mathcal K(\exp(j\omega))$ in (\ref{eq:K_exp})
over $[0,\pi]$ for different values of $\rho$. The colour coding is
as for Figure~\ref{fig:K_exp1}.} \label{fig:K_exp2}
\end{figure}

Now when values are independent of the spatial index $i$, equation
(\ref{alg_var_expd}) yields {\footnotesize
\begin{align}
y_i(k+1)&=(1+2\rho)y_i(k)-(2\rho+\rho^2)y_i(k-1)+\rho^2y_i(k-2)\nonumber \\
&+\frac{1-\rho}{1+\rho}[x_i(k+1)-x_i(k)+-\rho^2(x_i(k-1)-x_i(k-2))]
\end{align}
} The transfer function linking the measured to consensus variables
is then {\small
\begin{align} \label{eq:K_exp}
\mathcal
K(e^{j\omega})&=\frac{\frac{1-\rho}{1+\rho}[1-e^{-j\omega}-\rho^2(e^{2j\omega}-e^{3j\omega})]}
{1-(1+2\rho)e^{-j\omega}+(2\rho+\rho^2)e^{-2j\omega}-\rho^2e^{-3j\omega}}\nonumber \\
&=\frac{\frac{1-\rho}{1+\rho}[1-\rho^2e^{-2j\omega}]}{1-2\rho
e^{-j\omega}+\rho^2e^{-2j\omega}}\nonumber \\
&=\frac{\frac{1-\rho}{1+\rho}[1-\rho^2e^{-2j\omega}]}{(1-\rho
e^{-j\omega})^2}
\end{align}}
Evidently, the transfer functions $\mathcal K(e^{j\omega})$ and
$\mathcal H(e^{j\omega})$ in (\ref{eq:H_exp}) are not that different
in terms of the way their magnitude depends on $\omega$ and $\rho$.
Indeed, once again one can verify that if $1-\rho$ is small and
$\omega=1-\rho$, then $\mathcal K$ is approximately $1/2$. So the
spatial and temporal bandwidths are about the same. This appears
consistent with the assumption that a spatial progression of one hop
occurs in each time update, i.e. values propagate with effectively
unit velocity. Of course, the poles and zeros for the spatial
transfer function lie symmetrically inside and outside the unit
circle, in contrast to the time-based frequency response. We display
the behaviour of $\mathcal K(e^{j\omega})$ near the origin and over
$[0,\pi]$ respectively in Figures~\ref{fig:K_exp1} and
\ref{fig:K_exp2}.

\begin{figure}[!t]
\centering
\includegraphics[width=0.53\textwidth]{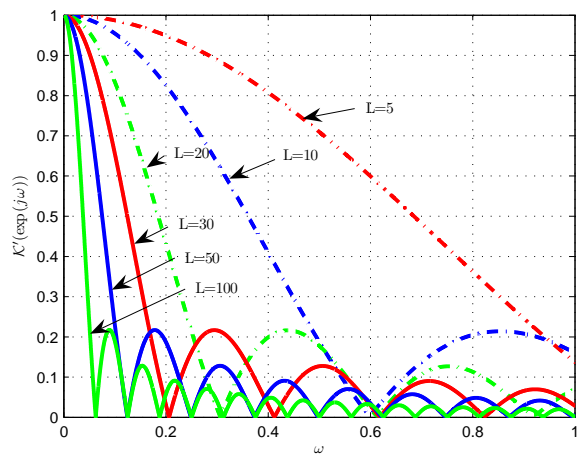}
\caption{Plot of $\mathcal K'(\exp(j\omega))$ in (\ref{eq:K_win})
for different values of $L$.} \label{fig:K_win}
\end{figure}

The treatment of time variation when the uniform finite window
approach is being used is also simple. Analogously to
(\ref{eq:K_exp}), we can obtain for {\small
\begin{align} \label{eq:K_win}
\mathcal K'(e^{j\omega})
=\frac{1}{2L+1}[1+2(e^{-j\omega}+e^{-2j\omega}+\cdots+e^{-Lj\omega})]
\end{align}}
Figure~\ref{fig:K_win} shows plots of this expression for different
values of $L$. When $\frac{2}{L+1/2}$ is small (this corresponds to
the condition $1-\rho$ is small for the exponential weighting case),
we can see that the frequency at which $|\mathcal K'(e^{j\omega})|$
assumes the value $1/2$ is approximately $\frac{4}{L+1/2}$.

Finally, we remark that the considerations applicable to spatial
variation without temporal variation or to temporal variation
without spatial variation will apply (because of the linearity of
the whole system) to a situation where both types of variation are
present in the measured variables. Thus if the measured variable
variation places them in the spatial bandwidth and outside the
temporal bandwidth, or the reverse, the consensus averaging process
will attenuate or suppress the variation.

%
%

%
\section{Conclusions} \label{Sec_Concl}

We have studied local average consensus in distributed measurement
of a variable using 1D sensor networks.  Distributed local consensus
algorithms have been designed to address first the case where the
measured variable has spatial variation but is constant in time, and
then the case where the measured variable has both spatial and
temporal variations. Two schemes for local average computation have
been employed: exponential weighting and uniform finite window.
Further, we have analyzed temporal-spatial frequency response and
noise propagation associated to the algorithms. Arbitrary updating
weights and random spacing between sensors have been analyzed in the
proposed algorithms.

In work which has yet to be submitted for publication, we have
studied two dimensional arrays. With a uniform grid, results rather
like those with fixed $\rho$ and $L$ can be obtained, but for a
general two dimensional array, a theory appears needed and is
currently under development.

%
\IEEEpeerreviewmaketitle

%
%
%
%
%
\section{ACKNOWLEDGMENTS}

This research is supported by ARC Discovery projects DP110100538 and
DP120102030. National ICT Australia (NICTA) is funded by the
Australian Government as represented by the Department of Broadband,
Communications and the Digital Economy and the Australian Research
Council through the ICT Centre of Excellence program.

\ifCLASSOPTIONcaptionsoff
  \newpage
\fi



\bibliographystyle{IEEEtran}
\bibliography{consensus}
\end{document}